\documentclass[epj]{svjour}
\usepackage{graphics}
\begin{document}
\title{Magnetic Moments of Baryons containing all heavy quarks in Quark-Diquark Model}
\author{Kaushal Thakkar\inst{1,*},   Ajay Majethiya\inst{2} \and P. C. Vinodkumar\inst{3}
}                     
%
%
\institute{Department of Applied Sciences \& Humanities, GIDC Degree Engineering College, Abrama, Navsari-396406, India \and
V S Patel College of Arts and Science, Bilimora-396321, India \and Department of Physics, Sardar Patel University, Vallabh Vidyanagar-388 120, India \\
$^*$Corresponding Author: kaushal2physics@gmail.com}
\date{Received: date / Revised version: date}
%
\abstract{
The triply heavy flavour baryons are studied using the Quark-diquark description of the three-body system. The confinement potential
for present study of triply heavy flavour baryons is assumed as coulomb plus power potential with power index $\nu$. We have solved
Schrodinger equation numerically to calculate the masses of triply heavy flavour baryons. The masses and magnetic moments of triply
heavy flavour baryons are computed for different power indices, $\nu$, starting from 0.4 to 1.0. The predicted masses and magnetic
moments are in good agreement with other theoretical predictions.
} 
\maketitle
\section{Introduction}
The field of heavy baryon spectroscopy is experiencing a rapid advancement in recent times. Numerous experimental data on heavy flavour baryons are reported. Particle Data Group (PDG 2014) has listed ground states of  heavy baryons with single heavy quark \cite{pdg14}. Though theoretical data on the properties of heavy flavour baryons are available in literature, the ground state masses of  doubly and triply heavy flavour baryons have not been measured experimentally. The SELEX Collaboration have claimed the observation of a $\Xi_{CC}^+$ double charm baryon through the decay channel from all possible doubly heavy flavour baryon states \cite{{SELEX},{SELEX5},{SELEX2005}}.
Thus the theoretical predictions of the doubly and triply heavy baryon
mass spectrum have become a subject of renewed interest and are relevant  due to the recent experimental
efforts at CMS, SELEX, LHCb etc. These experimental groups
have been successful in discovering heavy baryonic states along with other heavy
flavour mesonic states and it is expected that more heavy flavour baryon states will
be detected at J-PARC, PANDA, LHCb, and Belle II in their future efforts.

As in the case of $\Delta^{++}$ (uuu), $\Delta^{-}$ (ddd) and $\Omega^{-}$ (sss) baryons in the light flavour sector, QCD predicts similar states made up of charm quarks, $\Omega_{ccc}^{++}$ (ccc) and bottom quarks, $\Omega_{bbb}^{-}$ (bbb) in the heavy flavour sector. Though such a state yet to be observed experimentally. After the observation of the doubly charmed baryon by the SELEX group, it is expected that the triply heavy flavour baryonic
state may be in the offing very soon.
Though considerable amount of data on the properties of the singly-heavy
baryons are available in literature \cite{Kaushal16,Kaushal13,Ajay08,D.Ebert,Chen15,Yoshida15,Bijker,Glozman,Loring}, only sparse attention has been
paid to the spectroscopy of double and triple-heavy flavour baryons, perhaps
due to the lack of experimental incentives.
Theoretically, baryons are not only the interesting systems to study the quark
dynamics and their properties but are also interesting from the point of view of simple systems to study three body problems. Out of many approaches and
methods such as the QCD sum rules, QCD bag model, the effective field theory, lattice QCD, hyper central model, potential model etc., available for the three body systems \cite{Aliev14,Padmanath,Aliev13,Stefan,Felipe,Yu,Vijande,Bhavin}, we employ here the Quark-diquark approach to
study the triply heavy flavour baryons.  The magnetic moments of heavy
flavour baryons are computed based on the non-relativistic quark model using
the spin-flavour wave functions of the constituting quarks. The binding energy
effects are considered by defining an effective mass of the bound quarks within
the baryon for computing the magnetic moments. In the present study, we
compute the effective masses of the constituent quarks within the baryon with
different combinations of heavy flavour quarks. We repeat our computations
by varying the power index $\nu$ of confinement potential  from 0.4 to 1.0 to have an
idea about the most suitable form of inter quark potential that yields the static
properties of the triple heavy baryons.

The study presented in this paper is organized as follows: In section 2 the Quark-diquark model and a brief introduction of Coulomb plus power potential employed for the present study are described. The details of magnetic moments and its dependence of confinement through effective quark mass are presented in section
3. We analyse and discuss our results and make our own conclusions in section 4.

\section{Theoretical Framework: Quark-diquark model}

For the description of the 3-body system like the baryons, we have employed the quark-diquark model with interquark potential of the coulomb plus power potential form to study the masses and magnetic
moments of baryons containing three heavy flavour quarks.
Following Gell-Mann's suggestion of the possibility of quark-diquark structure for baryons \cite{M.
Gell-Mann1964}, various authors have introduced effective degrees of freedom of diquarks in order to
describe baryons as composed of a constituent diquark and quark \cite{M. Ida1966,D. B. Lichtenberg1967,M. Anselmino1993,F.Wilczek2005}. The presence of a coherent diquark
structure within baryons helps us to treat the problem of three-body interaction to that of two two-body interactions.\\

In this case, Hamiltonian of the baryon is expressed in terms of a
diquark Hamiltonian ($H_{jk}$) plus quark-diquark Hamiltonian ($H_{i,jk}$) as \cite{W. S. Carvalho1994,CJPH}
\begin{equation}\label{eq:1}
 H=H_{jk}+H_{i,jk}
\end{equation}
The internal motion of the diquark($jk$) is described by
\begin{equation}\label{eq:2}
H_{jk}=\frac{{p}^2}{2m_{jk}}+V_{jk}(r_{jk})
\end{equation}
 and the Hamiltonian
of the relative motion of the diquark($d$)- quark($i$) system is described by
\begin{equation}\label{eq:3}
H_{i,d}=H_{i,jk}=\frac{{q}^2}{2m_{i,jk}}+V_{i,jk}(r_{id})
\end{equation}
where, ($p, r_{jk}$) represent the relative momenta and coordinate of the quarks within the diquark system and ($q, r_{id}$) is the relative momenta and coordinate of the quark-diquark system. The reduced mass of the two body systems appeared in Eqn.(\ref{eq:2}) and Eqn.(\ref{eq:3}) respectively are defined as
\begin{equation} m_{jk}=\frac{m_{j} m_{k}}{m_{j}+m_{k}},\\ m_{i,jk}=\frac{m_{i}(m_{j}+ m_{k})}{m_{i}+m_{j}+m_{k}}
\end{equation}\label{eq:4}
For the present study, we have assumed colour coulomb plus power potential for the interquark potential of Eqn.(\ref{eq:2}) as well as for the quark-diquark interaction of Eqn.(\ref{eq:3}). \\Accordingly, the diquark potential can be written as,
\begin{equation}\label{eq:05}
V_{jk}=-\frac{2}{3}\alpha_s\frac{1}{r_{jk}}+ b{\,\,}  r^{\nu}_{jk}
\end{equation}
and the quark-diquark potential as
\begin{equation}
V_{i,jk}=-\frac{4}{3}\alpha_s\frac{1}{r_{id}}+ b{\,\,}  r^{\nu}_{id}
\end{equation}
where, $r_{id}$ is the quark-diquark separation distance, $\nu $ is the exponent
corresponding to the confining part of the potential and $b$ is the strength
of the potential, which is assumed to be same for the di-quark interaction as well as
between the quark-diquark interaction. The Schrodinger equation corresponds to the Hamiltonian given in Eqn.(\ref{eq:1}) is numerically solved using the Runge-Kutta method. The degeneracy of the states are removed by introducing the spin dependent
interaction potential given by \cite{S. S. Gershtein2000}
\begin{equation}
V^{(d)}_{SD}(r_{jk}) = \frac{2}{3}\alpha_s
\frac{1}{3 m_{j} m_{k}}{\bf S_{j} \ \cdot S_{k}} [4\pi \delta(r_{jk})]
\end{equation}
among the diquark states, and
\begin{equation}\label{eq:2.10}
V^{(i-d)}_{SD}(r)= \frac{4}{3}\alpha_s \frac{1}{3 m_{i} 2 m_{jk}}({\bf{S_{d}+ L_{d}) \ \cdot S_{q}}} [4\pi
\delta(r_{id})]
\end{equation}
among the quark-diquark($id$) system. \\

The model parameters are
fixed for the potential index $\nu=$ 0.4 to 1.0 to get the spin average mass
of the system $\Omega^{++}_{ccc}$ (4746), $\Omega^{+}_{ccb}$ (8021), $\Omega^{+}_{bbc}$(11283),
$\Omega^{-}_{bbb}$(14370). All other properties are predicted without changing any of these parameters.
The model parameters of  triple heavy flavour baryons are shown in Table \ref{tab:2.32}, \ref{tab:2.33} and \ref{tab:2.34}.
Our results for $J^P=\frac{1}{2}^+$ and $\frac{3}{2}^+$ are also compared with other theoretical model
predictions. It is interesting to note that our predictions are in good agreement with other theoretical model predictions.
The mass variations of the heavy baryons with respect to $\nu$ from 0.4 to 1.0 are found to
be around 100 Me$V$ only.\\

\begin{table}
\caption{Model parameters for $\Omega_{ccc}$ systems in
quark-diquark model.} \vspace{0.01in}\label{tab:2.32}
\begin{center}
\begin{tabular}{lcccc}
\hline\hline
$\alpha_{s}$&$\nu$ &$b$&$m_{c}$&\\
\hline
            &   0.4 &   0.1975  &  1.360  &   \\
            &   0.6 &   0.1010  &  1.435   &       \\
    0.20    &   0.8 &   0.054  &   1.480  &       \\
        &   1.0 &   0.033  &   1.500   &       \\
\hline\hline
\end{tabular}
\end{center}
\end{table}

\begin{table}
\caption{Model parameters for $\Omega_{ccb}$ and $\Omega_{bbc}$
systems in quark-diquark model.} \vspace{0.01in}\label{tab:2.33}
\begin{center}
\begin{tabular}{lcccc}
\hline\hline
$\alpha_{s}$&$\nu$ &$b$&$m_{c}$&$M_{b}$\\
\hline
            &   0.4 &   0.1875  &  1.360  & 4.720  \\
            &   0.6 &   0.0955  &  1.435   &  4.780     \\
    0.17     &   0.8 &   0.0495 &   1.480  &  4.815     \\
         &   1.0 &   0.0282  &   1.500   &  4.840     \\
\hline\hline
\end{tabular}
\end{center}
\end{table}

\begin{table}
\caption{Model parameters for $\Omega_{bbb}$ system in quark-diquark
model.} \vspace{0.01in}\label{tab:2.34}
\begin{center}
\begin{tabular}{lcccc}
\hline\hline
$\alpha_{s}$&$\nu$ &$b$&$m_{b}$&\\
\hline
0.17&   0.4 &   0.097   &  4.720     \\
&   0.6 &   0.015   &  4.780    \\
\hline\hline
\end{tabular}
\end{center}
\end{table}

\begin{table}
\begin{center}
\caption{Triply heavy baryon masses (in MeV) in quark-diquark
model}\label{tab:2.35}
\begin{tabular}{clllll}
\hline\hline
Baryon & Model&{$\textbf{J}^P=\frac{1}{2}^+$}&Others&{$\textbf{J}^P=\frac{3}{2}^+$}&Others\\
\hline
$\Omega^{++}_{ccc}$&$\nu=0.4$&$-$&$-$&4760&4965 {\cite{Roberts2007}}\\
&\ \ \ \ \ \ 0.6&$-$&$-$&4760&4760 {\cite{Yu}} \\
&\ \ \ \ \ \ 0.8&$-$&$-$&4760&4803 \cite{Martynenko2007} \\
&\ \ \ \ \ \ 1.0&$-$&$-$&4760&4790 {\cite{Bjorken85}} \\
&\ \ \ \ \ \ &$-$&$-$&  & 4773 {\cite{Migura2006}}\\
&&&&&    4777 {\cite{Bernotas2008}}\\
\\
$\Omega^{+}_{ccb} $&$\nu=0.4$&7999&8245 {\cite{Roberts2007}}& 8032&8265{\cite{Roberts2007}}\\
&\ \ \ \ \ \ 0.6&8002 &8018 \cite{Martynenko2007}& 8031&7980 {\cite{Yu}}\\
&\ \ \ \ \ \ 0.8&8004&$-$&8028&8025 \cite{Martynenko2007}\\
&\ \ \ \ \ \ 1.0&8005&7984 {\cite{Bernotas2008}}&8027&8200 {\cite{Bjorken85}} \\
&\ \ \ \ \ \ &&&&8005 {\cite{Bernotas2008}}\\
\\
$\Omega^{0}_{bbc} $&
$\nu=0.4$&11274&11535 {\cite{Roberts2007}}&11287&11554 {\cite{Roberts2007}}\\
&\ \ \ \ \ \ 0.6&11275&11280 \cite{Martynenko2007}& 11286&11190 {\cite{Yu}} \\
&\ \ \ \ \ \ 0.8& 11276&& 11285&11287 \cite{Martynenko2007}\\
&\ \ \ \ \ \ 1.0& 11277&11139 {\cite{Bernotas2008}}&11284 &11480 {\cite{Bjorken85}}\\
&\ \ \ \ \ \ & && &11163 {\cite{Bernotas2008}}\\
\\
$\Omega^{-}_{bbb} $&$\nu=0.4$&$-$&$-$&14370&14834 {\cite{Roberts2007}}\\
&\ \ \ \ \ \ 0.6&$-$&$-$&14370&14370 {\cite{Yu}}\\
&\ \ \ \ \ \ 0.8&$-$&$-$&14370&14569 \cite{Martynenko2007}\\
&\ \ \ \ \ \ 1.0&$-$&$-$&14370&14760 {\cite{Bjorken85}}\\
&\ \ \ \ \ \ &$-$&$-$&&14276 {\cite{Bernotas2008}}\\
\hline\hline
\end{tabular}
\end{center}
\end{table}

\section{Magnetic moments of  QQQ baryons}
According to the idea of constituent quark mass \cite{zeldovich 1967}, the hadron mass M is given by the sum of the masses of its constituent quarks. Accordingly, the kinetic and binding energies must be incorporated in the constituent quark mass. We define an effective mass to the constituting quarks within the baryons that takes care of the bound state effects including its internal motions and interactions among quarks as \cite{Bhavin2008,Thakkar}

\begin{equation}
m^{eff}_{i}=m_i\left( 1+\frac{\langle H\rangle}{\sum\limits_{i}m_{i}}\right)
 \end{equation}

For the computation of the magnetic moments, we consider the mass
of bound quarks inside the baryons as its effective mass taking in
to account of its binding interactions with other two quarks
described by the Hamiltonian given in Eqn.(\ref{eq:1}). The
effective mass for each of the constituting quark  $m^{eff}_{i}$
as defined  in Eqn.(9), where, $\left<H\right>=E+\left<V_{spin}\right>$ are computed such that the corresponding mass of the baryon with
spin angular momentum, J is given by
\begin{equation}\label{eq:2.14}
M_B^{J}=\sum\limits_{i}m_{i}+\left<H\right>_{J}=\sum\limits_{i}m^{eff}_{i}
\end{equation}
Accordingly, the effective mass of the $c$ and $b$ quarks will be different when it is in the
baryonic states of ccb combinations or bbc combinations
as $\left<H\right>_{ccb}\neq\left<H\right>_{bbc}$. The effective masses of the constituting quarks in the case of
the typical baryonic states  are listed in the Table \ref{tab:6}.\\

The magnetic moment of baryons are obtained in terms of the spin, charge and effective mass of the
bound quarks as \cite{{Ajay08},Bhavin2008}
\begin{equation}\label{eq:2.15}
\mu_B=\sum\limits_{i}\left<\phi_{sf}\mid\mu_{iz}\mid\phi_{sf}\right>
\end{equation}
where
\begin{equation}\label{eq:2.16}
\mu_{iz}=\frac{e_{i} \sigma_{iz}}{2m_{i}^{eff}}
\end{equation}
Here, $e_{i}$ and $\sigma_{iz}$ represents the charge and the spin
projection of the quark constituting the baryonic state. Using spin flavour wave function and effective masses,
we compute the magnetic moments of the triple heavy flavour baryons.\\
\par The spin flavour wave functions and the magnetic moments in terms of their constituent quark magnetic moments
corresponds to the various combinations of b, c quarks composition are listed in Table \ref{tab:3.20}. Using the
ground state masses predicted for the triple heavy flavour (QQQ, Q $\in$ c, b) baryons based on Quark-diquark approaches
 as discussed in section 2, we compute the effective masses of the constituting heavy quarks.  The predicted magnetic moments of all the spin
$J^{P}=\frac{1} {2}^{+}$ and spin $J^{P}=\frac{3} {2}^{+}$ triple heavy baryons based on quark-diquark model are listed in Table (\ref{tab:7}).
\begin{table}
\begin{center}
\caption{Spin-flavour wave functions and magnetic moments of
triple heavy flavour baryons.} \label{tab:3.20}
\begin{tabular}{lllllll}
\hline\hline
Baryon &Spin-flavour wave function&Megnetic moment\\
\hline
$\Omega^{*++}_{ccc}$&$c_{+}c_{+}c_{+}$&$3 \mu_{c}$\\
$\Omega^{+}_{ccb}$&$\frac{\sqrt{2}}{6}(2b_{-}c_{+}c_{+}-
                                           c_{-}b_{+}c_{+}-
                                           b_{+}c_{-}c_{+}+
                                           2c_{+}b_{-}c_{+}-
                                           c_{+}c_{-}b_{+}$\\
& \ \ \ \                                $-c_{-}c_{+}b_{+}-
                                           c_{+}b_{+}c_{-}-
                                          b_{+}c_{+}c_{-}+
                                           2c_{+}c_{+}b_{-})$&$\frac{4}{3}\mu_{c}-\frac{1}{3}\mu_{b}$\\
$\Omega^{*+}_{ccb}$&$\frac{1}{\sqrt{3}}(c_{+}c_{+}b_{+}+c_{+}b_{+}c_{+}+b_{+}c_{+}c_{+})$&$2\mu_{c}+\mu_{b}$\\

$\Omega^{0}_{bbc}$&$\frac{\sqrt{2}}{6}(2c_{-}b_{+}b_{+}-
                                           b_{-}c_{+}b_{+}-
                                           c_{+}b_{-}b_{+}+
                                           2b_{+}c_{-}b_{+}-
                                           b_{+}b_{-}c_{+}$\\
& \ \ \ \                                $-b_{-}b_{+}c_{+}-
                                           b_{+}c_{+}b_{-}-
                                           c_{+}b_{+}b_{-}+
                                           2b_{+}b_{+}c_{-})$&$\frac{4}{3}\mu_{b}-\frac{1}{3}\mu_{c}$\\
$\Omega^{*0}_{bbc}$&$\frac{1}{\sqrt{3}}(b_{+}b_{+}c_{+}+b_{+}c_{+}b_{+}+c_{+}b_{+}b_{+})$&$2\mu_{b}+\mu_{c}$\\
$\Omega^{*-}_{bbb}$&$b_{+}b_{+}b_{+}$&$3 \mu_{b}$\\
\hline\hline
(*indicates $J^P=\frac{3}{2}^+$ state.)
\end{tabular}
\end{center}
\end{table}

\begin{table}
\begin{center}\caption{Effective quark mass (in MeV) of
triple heavy flavour baryons.}
\label{tab:6}
\begin{tabular}{ccccc}
\hline \hline
Baryon  & $m_{c}^{eff}$& $m_{b}^{eff}$\\
\hline
$\Omega_{ccc}^{++}$ &    1586 & -            \\

\hline
$\Omega^{+}_{ccb}$  &  1531 &4936       \\

   \hline
$\Omega^{*+}_{ccb}$ &  1535  &   4951      \\

\hline
$\Omega_{bbc}$  &   1512   &4881        \\

   \hline
$\Omega^{*+}_{bbc}$   &   1514 & 4890        \\

\hline
$\Omega^{-}_{bbb}$    &  - & 4790     \\

\hline \hline
\end{tabular}

\end{center}
\end{table}

\begin{table}
\begin{center}
\caption{Magnetic moments of triply heavy baryons in quark-diquark model in units of Nuclear magneton $\mu_{N}$ }\label{tab:7}
\begin{tabular}{cccccccc}

\hline \hline
&\multicolumn{4}{c}{Potential index $\nu$}\\
\cline{2-5}
Baryon &0.4&0.6&0.8&1.0&RQM\cite{Faessler A2006}&NRQM\cite{Faessler A2006}&NRQM \cite{Silvestre1996}\\
\hline

$\Omega^{++}_{ccc}$ &1.182 &1.182 &1.182&1.182&$-$&$-$&1.023\\
\hline
$\Omega^{+}_{ccb}$
              &0.601 &0.576 &0.568&0.565&0.140&0.510&0.475\\
\hline
$\Omega^{*+}_{ccb}$&0.807 &0.768 &0.755&0.751&$-$&$-$&$-$\\
\hline
$\Omega^{+}_{bbc}$&-0.235 &-0.227 &-0.224&-0.223&-0.130&-0.200&-0.193\\
\hline
$\Omega^{*+}_{bbc}$&0.325 &0.297 &0.288&0.285&$-$&$-$&$-$\\
 \hline
$\Omega^{-}_{bbb}$ &-0.196 &-0.196 &-0.196&-0.196&$-$&$-$&-0.180\\
 \hline \hline
 (*indicates $J^P=\frac{3}{2}^+$ state.)
\end{tabular}
\end{center}
\end{table}

\section{Results and Discussions}
The masses and magnetic moments of triply heavy baryons in the Quark-diquark approach with coulomb plus power potential have been studied. It is interesting to note that our predictions of the mass of QQQ (Q $\in$ c, b) are in good agreement with existing predictions based on other
theoretical models.\\

The predictions of the magnetic moment of triply heavy flavour baryons studied
here are with no additional free parameters. Our results for magnetic moments
of triply heavy flavour baryons are listed in Table (\ref{tab:7}) and are compared with other
model predictions. The inter-quark interactions within the baryons are considered in the calculation of magnetic moments through the definition of effective
mass of the constituent quarks within the baryon (Eqn. 9).\\

Our predicted results for masses of $\Omega_{ccc}^+$ and $\Omega_{bbb}^-$ are in good agreement with the masses predicted by \cite{Yu}, while our predictions for $\Omega_{ccb}^+$ and $\Omega_{bbc}^0$ are in good agreement with masses predicted by \cite{Martynenko2007}. Our results for the magnetic moments of $\Omega_{ccc}^+$ and $\Omega_{bbb}^-$ are very close with NRQM predictions made by \cite{Silvestre1996}. Our results on the magnetic moments of $\Omega_{ccb}^+$ and $\Omega_{bbc}^+$ are in agreement with the  magnetic moments predicted by \cite{Faessler A2006}.\\

It is interesting to note that the masses and magnetic moments predicted in our model do not vary appreciably
with different choices of $\nu$ running from 0.4 to 1.0 as seen from Table (4) and Table (7). We look forward to see experimental data on these properties of triply heavy flavour baryons from future experimental facilities.

\end{document}